# Appearance of ferromagnetism property for Si nano-polycrystalline body and vanishing of electrical resistances at local high frequencies

Taku Saiki[1], Yukio Iida[1], Mituru Inada[1]

[1]*Faculty of Engineering Science, Kansai University, Japan*

*(3-3-35 Yamate, Suita, Osaka 564-8680, JAPAN)*

*\*Correspondence author*

*E-mail: tsaiki@kansai-u.ac.jp*

**Abstract**

Reduction in the skin effect for the sintered Si nanopolycrystalline body as an electricity conductor at a high frequency due to its nano-structure was studied. Singular vanishing of electrical resistances near a local high magnetic harmonic frequency of a few MHz was observed. This phenomenon has not been observed for conventional ferromagnetic metals. The measured electrical resistances changed to almost 0 mΩ at room temperature. At the same time, negative resistance of the sintered Si nano-polycrystalline body was observed. It will be applicable to electronic transmittance lines or semiconductors. Numerical calculation was also performed on the electrical resistance with frequency dependency while considering the electric field and magnetic field in the sintered Si nanopolycrystalline body. The experimental and calculated results are compared. The calculation could explain the variation of the relative permittivity of the Si nanopolycrystalline and the phenomenon for vanishing the resistivity at frequency of MHz theoretically.

Reduced Si nanoparticles from $SiO_2$ powder were synthesized by laser ablation in liquid. A Si nano-polycrystalline body made of the reduced Si nanoparticles was fabricated. It was found by measuring the magnetization property of the body that the sintered Si nano-polycrystalline body has ferromagnetism. Dangling bonds (unpaired electrons) have long been known to occur due to defects in Si crystals. Perfect Si without defective crystals has no dangling bonds. However, Si nanoparticles have many dangling bonds. High-density dangling bonds cause the sintered Si nanopolycrystalline to have ferromagnetism. In this study, the density of the unpaired electrons in the sintered Si nanopolycrystalline was observed using ESR. It has been clarified that the Si nanopowder and the sintered Si nanopolycrystalline have numerous dangling bonds. Both densities of the dangling bonds were evaluated.

**Introduction**

Reduction of the AC resistance owing to the skin effect at high frequency from MHz to GHz in the electronics devices or transmittance lines has been attracted now for realizing low-power-consumption. We can expect low-loss electric power lines and low-power-consumption electronic devices with low heat generation to be achieved by developing metals and semiconductors that have the reduced skin effect[1-4]. Research has also been conducted on the reductions in the skin effect for electric power lines using litz wire[2] or the negative permittivity of magnetic materials [5,6]. A multi-layered material, which consists of different material, such as metal and magnetic, is used for reducing the skin effect when a high-frequency current is conducted. This makes the effective permittivity equal zero when magnetic resonance occurs in magnetic material, negative permittivity at the real part of the relative permeability appears, and magnetic flux is canceled in multi-layered materials [6].

At the same time, a technology have been developed to make electrical circuits at low temperatures of below 470K using paste or ink consisting of Au[7], Ag[8-9], Cu[10], Fe[11,12], Al[11,12], Mg[12] nanoparticles owing to the phenomena of melting point depression. The metal nanoparticle paste and ink have been often sintered by finance or pulsed light like Xe flash lamp [8-9] or laser [12,13] in a short time. A free-form electrical circuit can be made in a short time and for low cost by using them. It is possible to make solids with maintaining or improving the property of the nanoparticles by sintering.

Moreover, many methods to produce metal or nonmetal nanoparticles exist. All of them, a method for producing the metal nanoparticles using laser ablation in liquids with high speed and low cost has especially attracted attention [11-19]. Many kinds of magnetic nanoparticles[22-34] (such as Fe[22-26], iron oxides[28], Au[29], Ni[28], and Pd[29-31] nanoparticles) have already been researched. Ferromagnetism materials will be applicable to high-frequency core inductors[33-34]. However, to realize our goal, only some materials meet the following conditions for the energy transfer or the semiconductor: 1) low resistivity, 2) low cost, and 3) abundance. Thus, we choose Si, which consists of a single atom and is used in standard semiconductors and solar panels. Some applications of Si are semiconductors and solar panels. Additionally, Si nanoparticles, which can emit visible light owing to the quantum size effects, can be used in color displays [15-21]. Also, researches on the spin wave propagation in Si nanoparticles[35-36], optical switch for optical computing using carrier effect by controlling the transparency of the material[37-40], application to air cells [13] are conducted now.

Unpaired electrons, i.e., dangling bonds, due to impurities in Si substrates have already been studied by many researchers. E ' and Pb defects due to surface oxidation of Si have already been confirmed [41-52]. Also, a theoretical research reports that ferromagnetism of the Si material is generated by interaction between dangling bonds that become unpaired electrons [35]. However, the ferromagnetism of bulk Si has not been observed experimentally until now. High density unpaired electrons are needed to generate ferromagnetism. Similar to Si material with crystal defects, the sintered Al polycrystalline bodies have been proven to have ferromagnetism even though common Al bulk has non-magnetism[32]. Common non-magnetic metals have been proposed to be able to change to being ferro-magnetic when we make solids from nanoparticles.

The goal of our research is to develop low-power-consuming semiconductor devices by applying the prepared Si nanopolycrystalline with a reduced skin effect to semiconductors. We have recently fabricated nano-structured metals by the bottom-up process [11-13]. A Sintered metal nano-polycrystalline body with low resistivity is a new material, and the resistivity has not been adequately investigated.

In this study, we investigated if sintered Si nanopaste is a ferro-magnetic material. We also report on a significant reduction in the resistance at a specific frequency that was also observed.

## Result
### Measurement by SEM

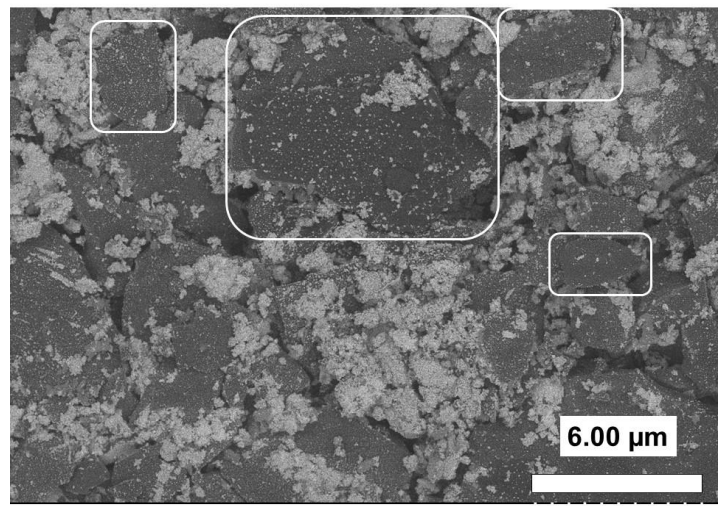

**Fig 1. (a) Reflection electron image obtained by SEM (Magnification: 5000 times).**

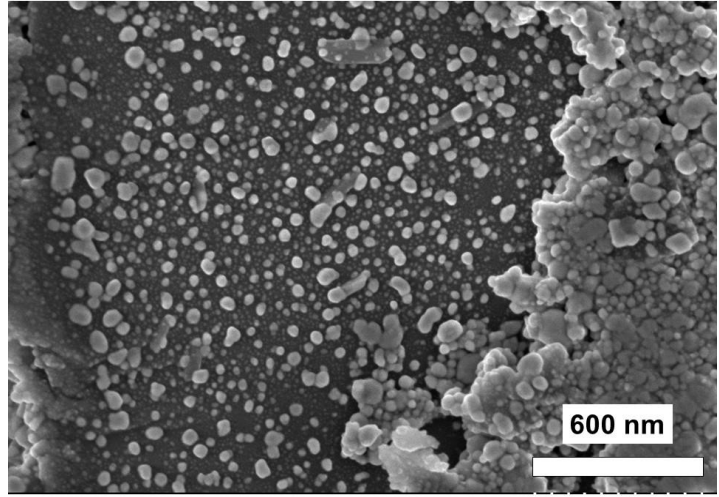

**Fig 1. (b) Secondary electron image obtained by SEM (Magnification: 50000 times)**

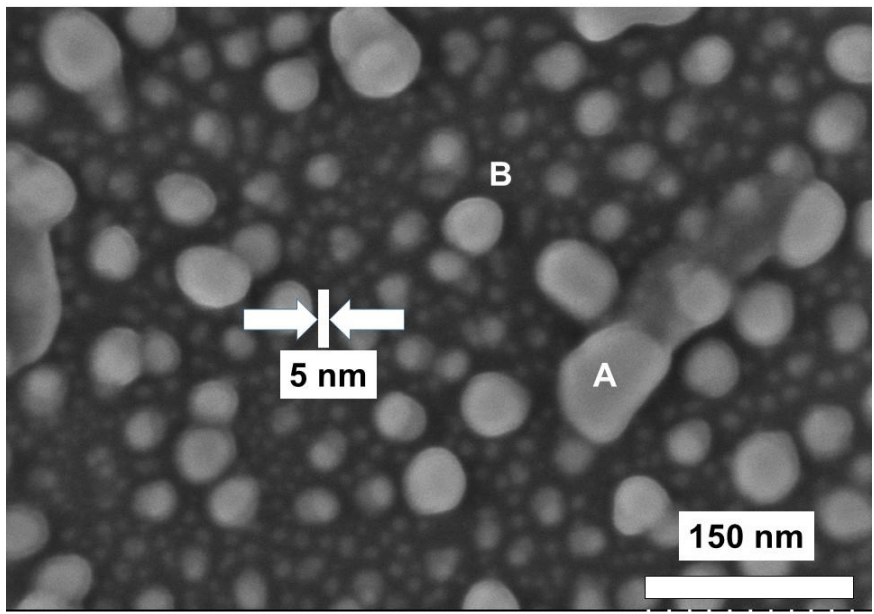

**Fig 1 (c) Secondary electron image obtained by SEM (Magnification: 200000 times)**

Once, we observed the structure of the sintered Si nano-polycrystalline body experimentally. A SEM image of a sintered Si nanopolycrystalline body is shown in Fig 1. It has been found by EDX analysis that the Si secondary particles consist of Si atoms. The mixed paste contains 6% of Ag atoms for all the atoms when we make Si nanopaste. After sintering the amount will reduce to a few % because the Ag concentrated on the surface should be removed. It is hard to know the concentration. Numerous Si secondary particles below 10 μm were observed as shown in white lines of Fig 1(a). The part surrounded by the white line is a secondary Si particle. The magnified SEM image in Fig 1(b) shows that this secondary particle is clearly composed of many minute Si nanoparticles. The expanded SEM image in Fig 1(c) shows that the Si nanoparticles are around a few nm although there is some variation in size. Fig 2 shows the results of EDX analysis by specific X-rays. Fig 2(a) and 2(b) show results for part A in black and part B in white. Si, O, and Ag atoms on the nanoparticles were measured at part A by specific X-rays in Fig 2(a), and Si and O atoms on the nanoparticles were measured at part B by specific X-rays in Fig 2(b). Although oxygen appears strongly in the specific X-ray spectrum, it seems to be particularly attached to the surface of the Si nanoparticle and constitute the interface of the secondary particle. Because the oxygen film is extremely thin, the resistance does not increase. The diameter of the Si nanocrystals should not change during sintering. The oxygen film is thin and does not increase the resistance. It is clear from Fig 1(c) that secondary Si particles consist of numerous nanocrystals with a mean diameter of below 2 nm in three dimensions. The Si nano-polycrystalline body emits a photoluminescence with a peak wavelength of 550 nm excited by a violet light source due to the small size of the Si nanocrystals [20,21].

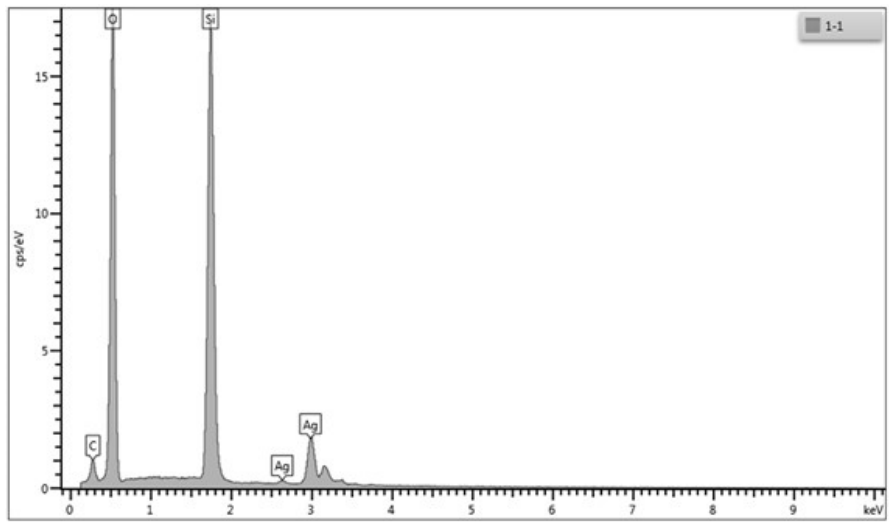

(a)

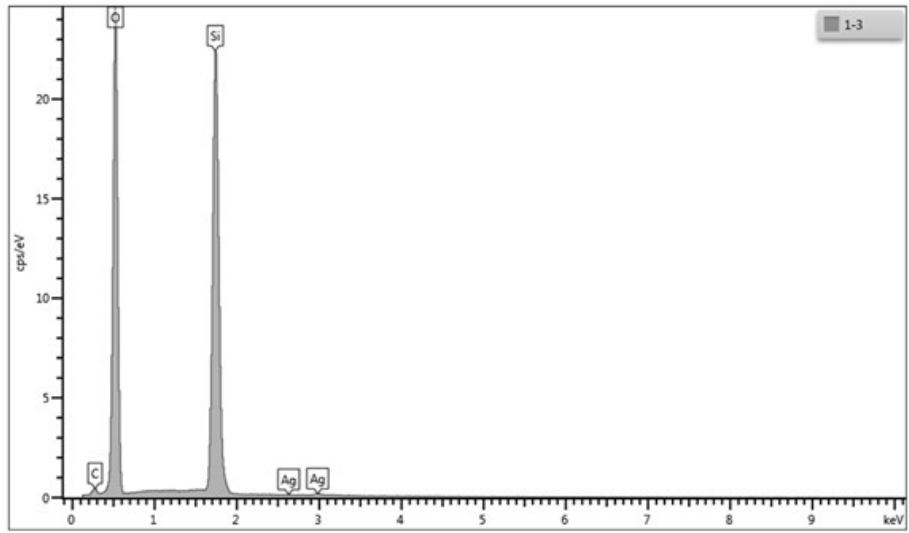

(b)

Fig 2. Results of EDX analysis by specific X-rays. (a) Part A in Fig 1(c), (b) Part B in Fig 1(c).

**Magnetization property**

Normal Si has an anti-magnetic property. The magnetization property of the reduced Si nanoparticles and the sintered Si nanopolycrystalline body was measured by VSM at a room temperature (293K) to determine if the sintered Si nano-polycrystalline body has ferromagnetism. The measured magnetization of reduced Si nanopowder and Sintered Si are shown as Fig 3(a). Also, Arrott plot of them is shown in Fig 3 (b). Magnetic parameters of them are shown in Table.1

The measured maximum magnetization of the reduced Si nanoparticles was 0.011 emu/g, and the measured maximum magnetization of the sintered Si nano-polycrystalline body was 0.075 emu/g. The magnetization of the sintered Si nano-polycrystalline body is 7 times larger than that of the reduced Si nanoparticles. The maximum magnetization of the conventional Fe bulk was 218 emu/g. The evaluated coercive force of the sintered Si nano-polycrystalline body was 200 Oe. From the measured magnetization property, the sintered Si nano-polycrystalline body includes a ferromagnetic phase because of the residual magnetization and coercive force as shown in Figs 3(a) and 3 (b). Because the magnetization curve saturated at 1000 Oe for the reduced Si nanoparticles, there was no evident spontaneous magnetization, but there were many defects of Si atoms with magnetic moment in the material, which react with the outer magnetic field.

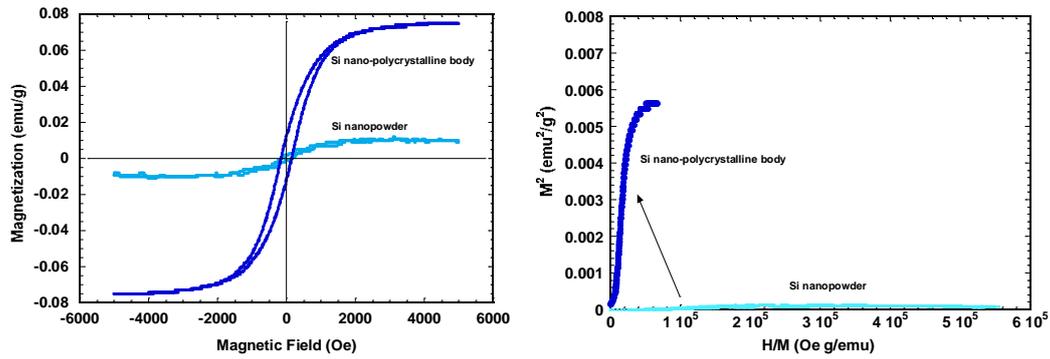

**Fig 3. Magnetic property. (a) Magnetization property. (b) Arrott plot.**

**Table 1. Magnetic parameter**

|  | Saturated Magnetization (emu/g) | Corrective force (Oe) |
|---|---|---|
| **Si nanopawder** | 0.011 | 200 |
| **Si nanocrystalline** | 0.08 | 200 |
| **Fe powder [23]** | 210 (mean size: 150nm) | 280 |
| **Fe oxide [27]** | 74 (mean size: 30nm) | 700 |
| **Ni doped ZnO [28]** | 4 | 90 |

**ESR analysis**

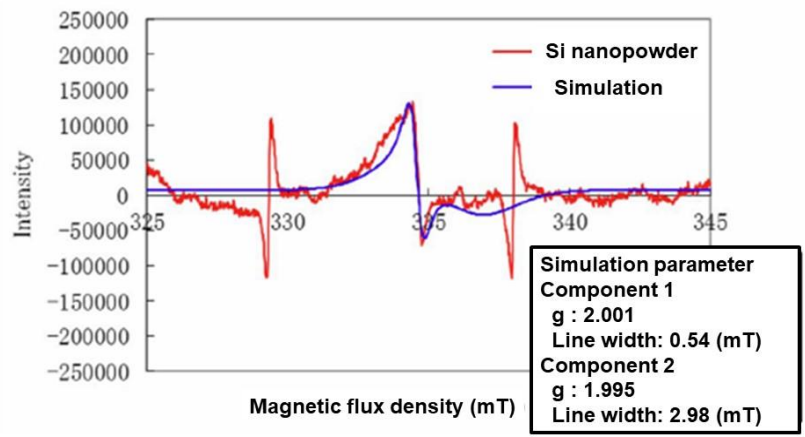

**Fig 4(a). Results of ESR analysis for Si nanopowder.**

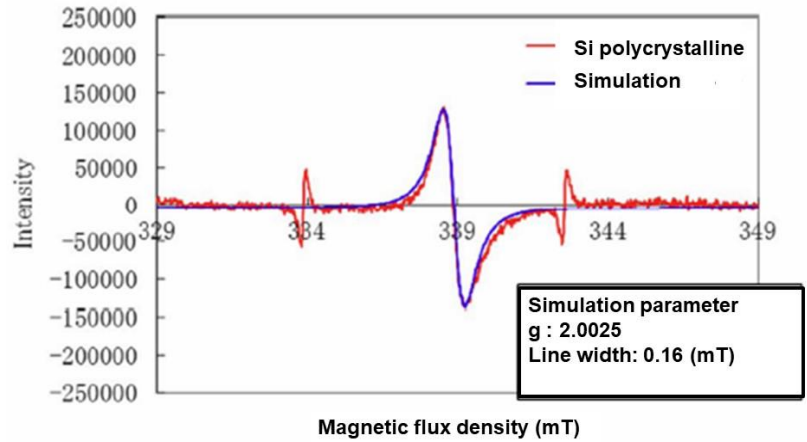

**Fig 4(b). Results of ESR analysis for sintered Si nanopolycrystalline.**

Densities of the dangling bond in the Si nanopowder and the Si nanopolycrystalline were measured by electron spin resonance (ESR) and then compared. Results of ESR analysis for the Si nanopowder and the sintered Si nanopolycrystalline are shown in Figs 4(a) and 4(b). The measured densities of the dangling bonds for the Si nanopowder and the Si nanopolycrystalline are shown in Table 2. The g for the Si nanopowder was 2.0000. It was considered from g that the Si nanopowder contains the dangling bonds of the Si oxides. Judging from the observed peak signal, the Si nanopowder contains more than two components. However, the structure of the Si nanoparticle is not recognized. It should be a kind of Si oxide. The g for the Si nanopolycrystalline was 2.0025, which was between those for the Si crystals (g=2.0050～2.0060) and that for the Si oxides (g=2.0000). It was found that intermediate dangling bonds between crystalline silicon and silicon oxide exist in sintered Si nanopolycrystalline. E 'centers and Pd centers in the reference [41-45] could not be separated in the electron spin resonance (ESR) spectrum. The amount of the dangling bonds calculated from the integrated value of the ESR spectrum and the area strength of the standard material (known amount of radical: DPPH) is estimated to be 1/2 that of the sintered Si nanopolycrystalline. Thus, it is concluded from this result that magnetization will amplified up to seven times for one of the Si nanopowders because of the interaction between unpaired electrons called as exchange interactions in the sintered Si nanopolycrystalline.

Table 2. Result on evaluated density of dangling bond.

|  | g value | Amount of dangling bond ($\times 10^{13}$ spin/g) |
|---|---|---|
| **Si nanopowder** | 2.0000 | 8.3 |
| **Si nanopolycrystalline** | 2.0025 | 16.0 |

**Measurement of resistivity**

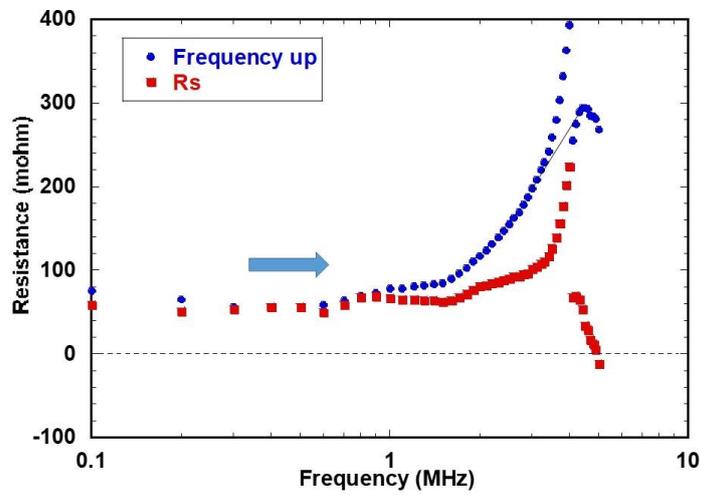

(a)

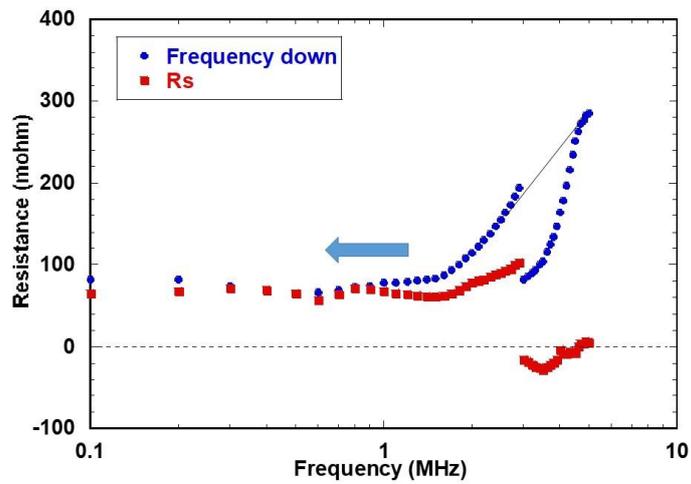

(b)

**Fig 5. Measured resistivity as function of frequency when current at 5MHz was 10mA.: (a)**

**increasing frequency, (b) decreasing frequency.**

Measured resistance of the sintered Si as a function of frequency is shown in Fig 5. The measured resistance with increasing frequency of the sine wave signal is shown in Fig 5(a), and that with decreasing frequency of the sine wave signal is shown in Fig 5(b). The blue line shows the measured resistance to measure including copper lines. The red line shows the resistance without the resistance of the copper lines. We observed a clear difference between the two results. Discontinuities of the resistances as a function of frequency were observed.

The resistance of the sintered Si nano-polycrystalline body increased with increasing frequency, and the resistance returned to 60 mΩ at 4 MHz. After that, the resistance decreased continuously, as shown in Fig 5(a). The resistance of the sintered Si nano-polycrystalline body remained close to zero with decreasing frequency, and the resistance decreased around -20 mΩ at 3 MHz. After that, the resistance decreased continuously from 100 mΩ as if the skin effect was weakened, as shown in Fig 5(b). The evaluated volume resistivity of the sintered Si nano-polycrystalline body was $1.2 \times 10^{-6}$ Ω·m at 100 Hz. A slight reduction in the resistance at 200 kHz was also observed.

The measured resistance of the sintered Si with the dependence on current is shown in Fig 6. The resistances were measured with decreasing frequency of the sine wave signal. The measured self-inductances of the sintered Si nano-polycrystalline body are shown in the inset figures of Figs 6(a), (b), and (c). Thus, from the current data, magnetic resonance occurs at these frequencies, and the relative permeability changes rapidly around these frequencies. The singular radical vanishing of the resistivity in the Si nano-polycrystalline body at high magnetic-resonance frequencies should occur because the sintered Si pastes are ferro-magnetic and exhibit negative permittivity around these frequencies. The self-inductance increased at 2.8 MHz, as shown in the inset of Fig 6(a). The self-inductance changed discontinuously at 200 kHz, 2.8 MHz, and 5 MHz. The resistance of the sintered Si increased to 400 mΩ at 2.8 MHz, as shown in Fig 6(a). At the same time, the current decreased to a half that at 100 kHz.
.

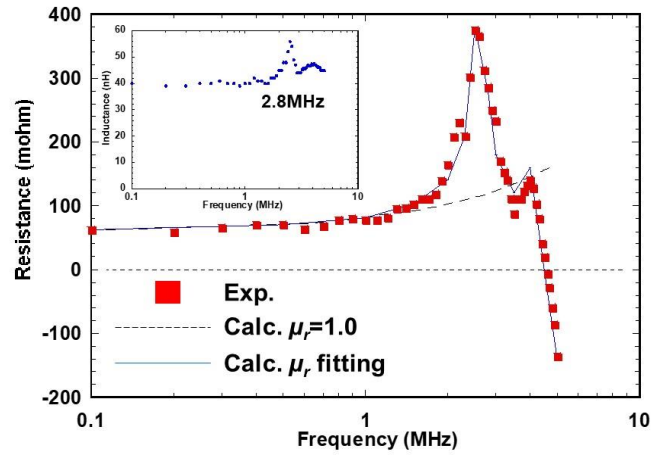

(a)

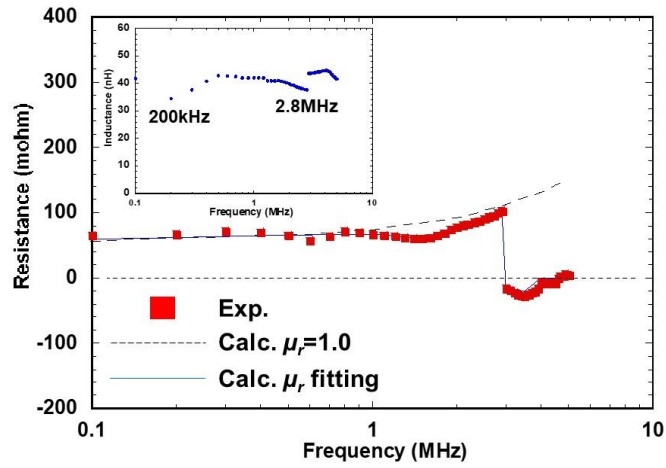

(b)

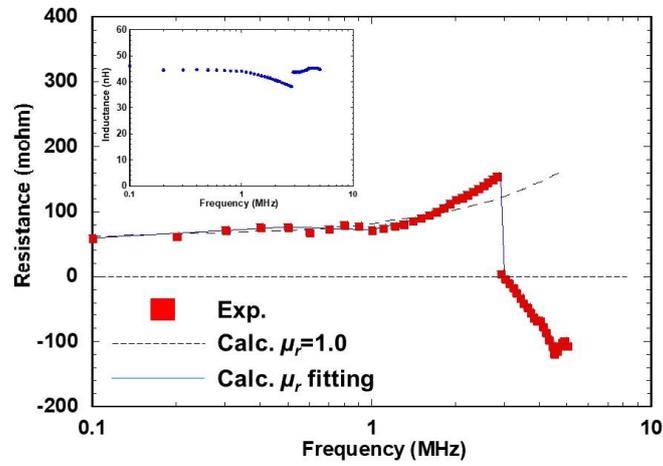

(c)

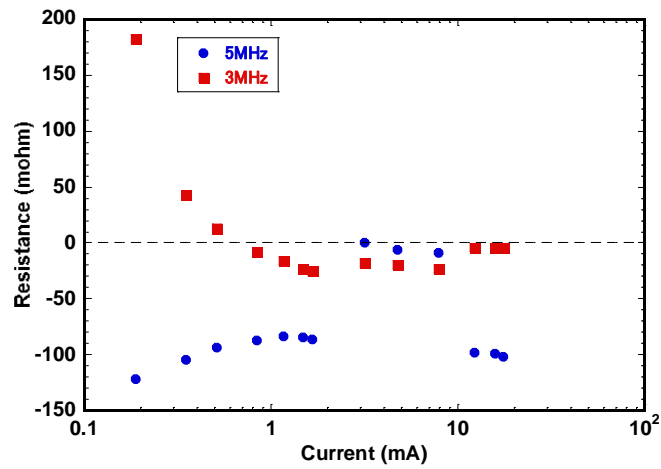

(d)

**Fig 6. Results on measured resistance as function of frequency. Currents at 5 MHz were (a) 0.17, (b) 10, and (c) 20 mA. (d) Resistances with dependence on current.**

The measured resistances of the Si nano-polycrystalline body with the dependence on current is shown in Fig 6(d). The resistance at 3 MHz was 190 mΩ when the current was 0.17 mA, and changed to 20 mΩ when the current was changed from 2 to 8 mA. Finally, the resistance was 0 mΩ asymptotically when the current was changed from 10 to 18 mA. The resistance at 5 MHz was -120 mΩ when the current was 0.17 mA and was 0 mΩ asymptotically when the current was changed from 3 to 8 mA.

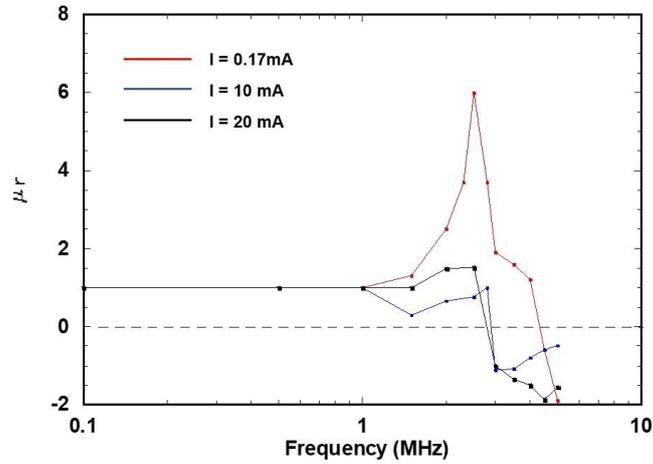

**Fig 7.　Evaluated relative permittivity as function of frequency. Currents at 5 MHz were (a) 0.17, (b) 10, and (c) 20 mA.**

The resistance at high frequency was calculated numerically while considering the permittivity of the Si nanopolycrystalline and the electric field and magnetic field. From the results, the resistance clearly varies in accordance with the relative permittivity of 1 as a function of the frequency as shown by the dashed lines in Fig 6(a), (b), and (c). This analysis result seems to be extraordinary because the sintered Si is metal. Thus, the relative permeability should degrade to below 1 as in the case of normal metal bulk. The lower relative permeability means an anti-magnetic property such as that appearing in conventional Al bulk. However, it was found from numerical calculation that the relative permeability of the sintered Si nano-polycrystalline body degraded to close to 1.

Thus, the relative permittivity was changed so as to fit the measured resistance. The evaluated relative permittivities as function of frequency are shown in Fig.7. The resistance was recalculated using the relative permittivity as shown by the blue solid line in Figs 6(a), (b), and (c). The resistance increased to 400 mohm when the frequency was 2.5 MHz as shown in Fig 6(a). This is because the generated magnetic field intensity is weak and the amplitude of the permittivity maximizes. It was thought that the phase of the permeability is inverted at 5 MHz, the real part of the relative permeability changed to be -1, and the direction of the electric field generated by the magnetic field is reversed.

For Fig 6(b), when the current increased, the intensity of the magnetic field increased by two orders of magnitude. The magnetic moment cannot keep up with time change of the magnetic strength owing to the high frequency, and at 3 MHz, the real part of the relative permeability is a binary change of 1 or -1. It was also found that the phase of the permeability was inverted at 1.5 MHz.

For Fig 6 (c), when the current increased, the magnetic field intensity became stronger. The magnetic moment cannot keep up with the temporal change of magnetic intensity owing to the high frequency, but at 3 MHz, the real part of the relative permeability is a binary change of 1.5 on the low frequency side and -1 on the high frequency side.

In the case in Fig 6 (a), the permeability was found to be consistent with that of the sintered Al having similar properties described in the reference [32], with the relative magnetic permeability at 2.5 MHz evaluated to be 7.

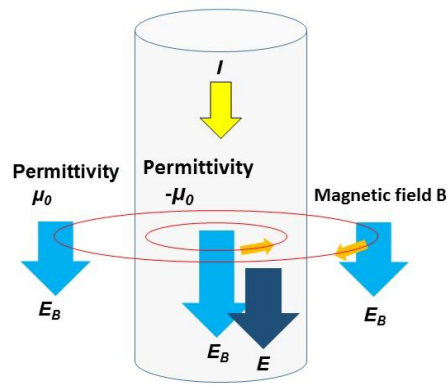

**Fig 8. Spatial distribution of permeability when magnetic resonance occurred and real part of permeability changed to negative at 3 MHz**

When the magnetic resonance occurred, the real part of the permeability changed to negative at 3 MHz, as show in Fig 8. The calculated results show that the real part of the permeability inside the sintered Si should be changed to $-\mu_0$. The term $E$ is the original electric field applied with the original sine wave signal, and $E_B$ is the electric field generated by the current induced magnetic field. These directions are the same because of the phase inversion of the $E_B$ when the real part of the relative permeability is negative. It was suggested that the electrons in the sintered Si nano-polycrystalline body were accelerated by the generated electric field.

However, at the same time, the sintered Si nano-polycrystalline body had a relative permeability of -1. The numerically calculated result show that the current does not flow linearly in micro scale of the sintered Si and generation of the eddy current is prevented due to the structure.

The resistance of a sintered Ag nano-polycrystalline body with non-ferromagnetism did not become zero at the magnetic resonance frequencies of a few MHz. Thus, this phenomenon is thought to be intrinsic for a metal nanopolycrystalline body with ferromagnetism.

Common metal has effective permeability below 1, such as 0.5, at the MHz level, and resistance become low due to the generated $E_B$. Thus, because the $E_B$ is weaker than that of metal in a ferro-magnetic metal nano-polycrystalline body, the resistance does not degrade to zero.

It is considered that the resistances of almost all ferro-magnetic metal nano-polycrystalline bodies changes to zero at frequencies of a few MHz with no connection to the relative permeability.

However, resistance from 3 to 20 mA changed to close to zero by some cause. The resistance also changed to negative around 5 MHz when the current was 20 mA, as shown in Fig 6(d). The cause seems to be that the magnetic momentum of unpaired electrons at the surface of Si nanocrystals cannot follow the varying magnetic field at a high frequency of 5 MHz.

In these experiments, we did not observe the reduction of the skin effect by which the current distribution in the Si is close to uniform due to the nano-structure. We believe that sintered Si nanopaste with a Si nano-polycrystalline body will be applicable to magnetic materials.

**Calculated magnetic field intensity and electric field intensity**

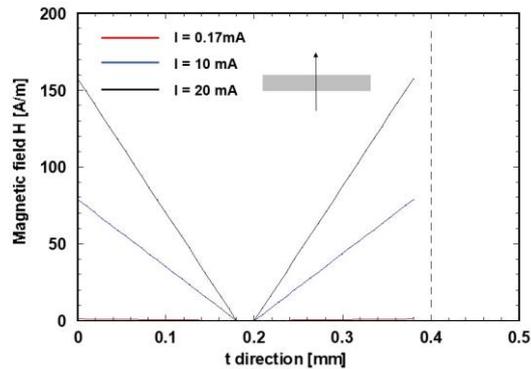

(a)

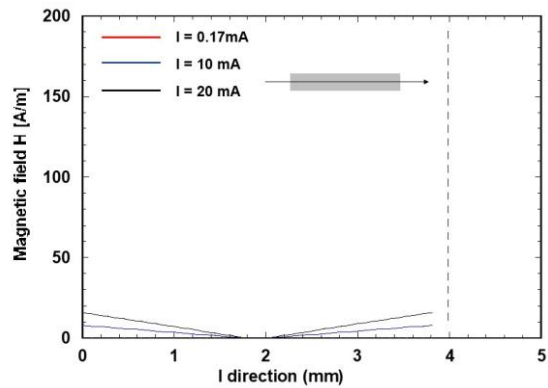

(b)

(a) Tick direction, (b) Length direction

Fig 9. Calculated Magnetic field intensity.

Fig 9 shows the calculated results for the distribution of the magnetic field intensity at the profile of the Si nanopolycrystalline. The current is shown as the effective value. The thin direction of the material is the thickness direction, and the long 4 mm direction is the length direction. The magnetic field intensity at the center of both directions was calculated. It was found that a strong magnetic field is mainly generated near the surface in the direction perpendicular to the thickness direction. Also, weak magnetic field of 1/10 was generated in the direction perpendicular to the length direction.

It was found that when the effective value of the current flowing in the Si nanopolycrystalline was 0.17, 10, and 20 mA, the calculated magnetic field intensity was 1, 80, and 160 A / m in the direction perpendicular to the thickness direction.

Fig 10 shows the calculated distribution of the electric field, in which the direction of the electric field is the same as the direction of the current and the intensity is proportional to current. Here, the electrical field intensity in the in the longitudinal direction was normalized by the AC electrical field intensity added by a power source. When the normalized electric field is negative, it means that the current barely flows in the forward direction. When the normalized electric field is positive and larger than 1, this means that the flow of current is assisted by the electric field and the resistance is reduced. For Fig 10 (a), when the current is 0.17 mA, the electric field intensity at the center continues to decrease from 0.1 to 4 MHz and is lowest at 2.5 MHz. The real part of the permeability is changed to be negative by magnetic resonance. The distribution of the electric field intensity was dented downward, and the distribution reversed from 2.5 to 5 MHz. For Fig 10 (b), when the current increased to 10 mA, the electric field intensity at the center continues to decrease from 0.1 to 2.9 MHz. The real part of the permeability is changed to be negative by magnetic resonance. The distribution of the electric field intensity was dented downward, and the distribution reversed from 3 to 5MHz.

For Fig 10 (c), when the current increased to 20 mA, the electric field intensity at the center continues to decrease from 0.1 to 2.9 MHz and is lowest at 2.9 MHz. The phase of the permeability is inverted to be negative by magnetic resonance. The distribution of the electric field intensity was dented downward, and the distribution reversed from 3 to 5 MHz.

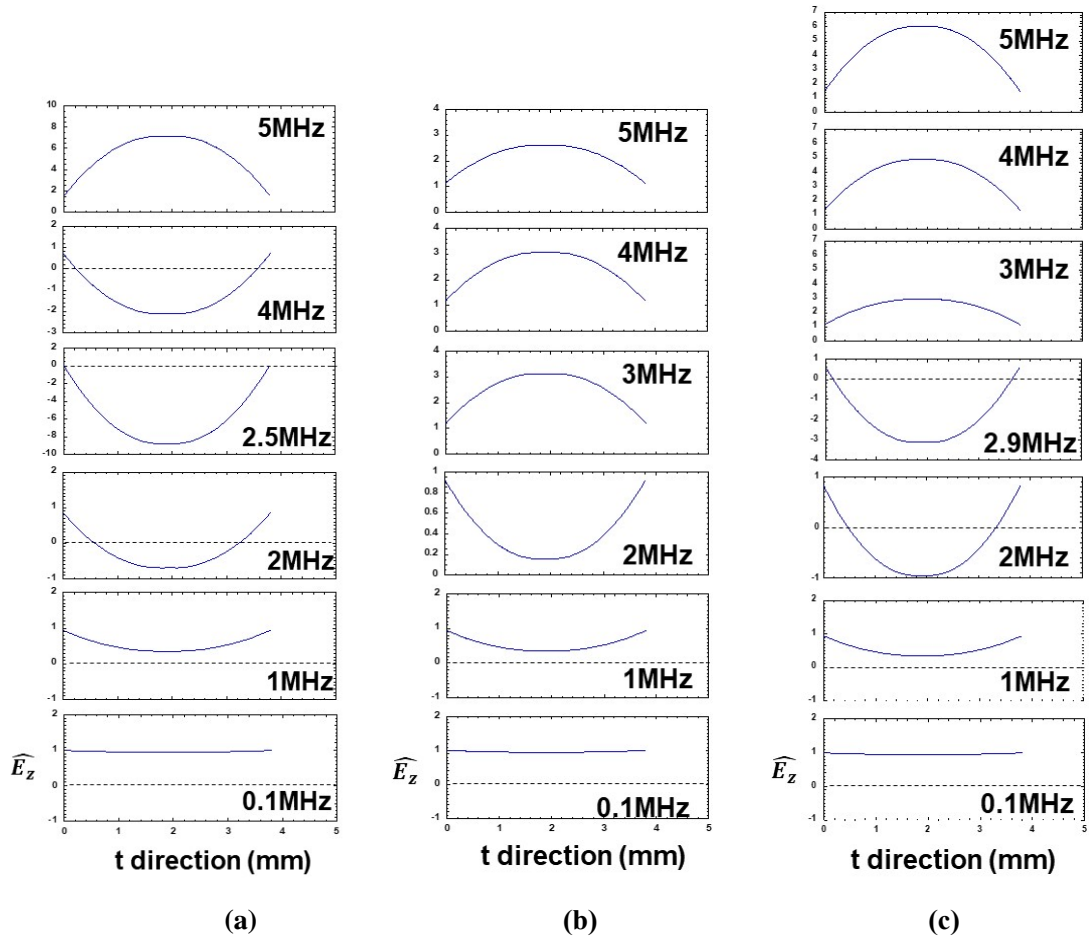

(a) (b) (c)

**Fig 10. Calculated electric field. (a)0.17mA, normarized by $E_{AC}$=0.0022V/m, (b)10 m A, normarized by $E_{AC}$=0.013V/m, (c) 20 mA, normarized by $E_{AC}$=0.026V/m.**

## Conclusion

Reduction in the skin effect for the sintered Si nanopolycrystalline for semiconductor material at a high frequency due to its nano-structure has been studied. Singular vanishing of electrical resistances near a local high magnetic harmonic frequency of a few MHz has been observed in the experiments. Negative resistance of the sintered Si nanopolycrystalline has been also observed. Numerical calculation has been also performed on the electrical resistance with frequency dependency while considering the electric field and magnetic field in the sintered Si nanopolycrystalline. The experimental and calculated results are compared. The calculation could theoretically explain the phenomenon for vanishing the resistivity at frequency of MHz. It was found by measuring the magnetization property that the sintered Si nanopolycrystalline have ferromagnetism. The density of the unpaired electrons in the sintered Si nanopolycrystalline was observed using ESR. It has been recognized that the sintered Si nanopolycrystalline have numerous dangling bonds.

## Experimental methods

### Production of materials

The laser ablation method in liquid is described in the paragraph below. When laser pulses are irradiated onto metal oxides in liquid, the metal oxides melt and resolve and the melted oxides are set outside the metal nanoparticles. The surrounding liquid cools the metal nanoparticles rapidly. The experimental setup for laser pulse ablation is shown here. A microchip Nd:YAG laser was used in this experiment. The maximum output averaged laser power was 250 mW, the laser wavelength was 1064 nm, the repetitive rate of the laser pulses was 18 kHz, and the pulse duration was 8 ns. A beam with a diameter of 6 mm ($1/e^2$) was focused using a lens with a focal length of 50 mm. Thus, the diameter of the focused beam was 20 μm at the front of each glass bottle. Glass bottle with a size of 38 mml x 20mm were used in the experiment. Reduced Si nanoparticles were produced by laser ablation in liquid. The liquid we used was pure water. The $SiO_2$ powder (mean diameter of 5 μm, purity 99.9%, Koujyund Chemincal, Japan) was used to produce reduced Si nanoparticles. The $SiO_2$ powder were mixed with the water in each glass bottle for the experiment. Glass bottle was set after the focused laser beam. The weight of the $SiO_2$ powder was measured using an electronic force balance. Their measured weight was 1.0 g. 12mL of pure water was placed in each glass bottle. Laser pulses were irradiated to the water with the metal oxide in glass bottle for 10 minutes. Here, we neglect the oxidation at the surface of metal nanoparticles. A magnetic stirrer was used to mix the liquid. The color after laser irradiation changed to gray, which is close to the color of the reduced Si powder. The powder after irradiating laser pulses in the water was dried to not change chemically.

A sintered Si nanopaste (Si nano-polycrystalline body) was made using the reduced Si nanoparticles. The dried Si nanopowders were mixed with 5mg of Ag nanopastes (NAG-10 Daiken Chemical); the viscosity of the paste was high. The size of the sintered Si nano-polycrystalline body was determined to be 4 x 10 x 0.3 mm. The current was conducted in the longitudinal dimension, and the resistivity at a high frequency was measured. The Si paste was sintered using an electrical hot plate (CHP-170AN, ASONE) at 473K(1 min.) and 533K (4 min), enabling us to obtain sintered Si pastes.

**Analysis of material and measurement method**

The magnetization properties of the reduced Si nanoparticles and the sintered Si nano-polycrystalline body were measured using a Vibrating Sample Magnetometer (VSM）(BHV-30T, Riken Denshi, Japan) at a room temperature (20°C). The sintered Si was observed by Scanning Electron Microscope (SEM）(S-4700 with low resolution and SU8240 with high resolution, Hitachi High-Technologies, Japan), and the existence of Ag, Si, and O atoms were analyzed by energy-dispersive X-ray spectrometry (EDX) (EMAX7000, Horiba, Japan).

We show the measurement condition for ESR analysis in Table. 3. FT-IR, S-FA200 type, JEOL was used for ESR analysis. The weight of the used sintered Si nano-polycrystalline body is 170mg. The size was 20 mm x 2.5mm x 1mmt.

Table. 3. Measurement condition

| | |
|---|---|
| **Microwave frequency** | 9.3-9.4 GHz |
| **Intensity of microwave** | 12mW |
| **Range of maneuvering magnetic field** | 20mT |
| **Modulated magnetic field frequency** | 100kHz |
| **Modulated magnetic field amplitude** | 1.0mT |
| **Sampling time** | 81.92ms |
| **Accumulation count** | 3 times |
| **Measurement temperature** | 10K (Liquid He cooling) |
| **Standard substance** | 1-diphenyl-2-picrylhydrazyl（DPPH） |
| **Index** | Mn marker ($Mn^{2+}$) |

The resistance and inductance of the sintered Si pastes were measured using an LCR meter (3532-50 LCR, high tester, Hioki, Japan). The inductance and resistance from 42 Hz to 5 MHz was measured. The inductors and electric power transmitters were assumed to have components of inductance and resistance. The phase angles of cascaded resistances and inductors in the stick-type sintered Si pastes were also measured.

**Calculation for skin effect**

We resolved the integration Maxwell-Faraday equation numerically and calculated the electric field generated by the current-generated magnetic field inside the metal. We observed spatial distribution of the electric field in the thickness and longitudinal direction. The electric field was added to the original electric field applied from the sine wave signal source. The resistance was calculated with the re-calculated electric field. We considered only the real part of the relative permeability. The special mesh was set to be 40 in both length and thickness directions. It was assumed that the permeability is uniform in the profile.

The current-generated magnetic field intensities in the directions of the length and thickness are given as

$$H_l(r) = \frac{Ir}{2Lt}$$
$$H_t(r) = \frac{IrL}{2t^3}$$

(1)

Where, $I$ is the AC currnent, the $L$ and t are the length and the thickness in the profile of the Si nano-polycrystalline body. The effective resistance at a high frequency is given as

$$R_{hef} = R_{DC}\left(1 + \frac{E_{Bav}}{E_{AC}}\right).$$

(2)

Where, $R_{DC}$ is the DC resistance, the $E_0$ is the original AC electric field intensity, and $E_{Bav}$ is the averaged electrical field intensity generated the current-generated magnetic field intensity. The resistance was calculated using the ratio of the electric field generated by the current-generated magnetic field to the original AC electric field. $R_{DC}$ was set to be 60 mohm.

**Additional Information**